\magnification 1200
\baselineskip 18 pt
%
\font\bigggfnt=cmr10 scaled \magstep 3
 2
\font\bigfnt=cmr10 scaled \magstep 1
%
\leftskip .25 in
\rightskip .25 in
\vglue .35 in
\def\Par{\par\vskip 3 pt}

\def\ea{{\it et al.\/}}
\def\ie{{\it i.e.}}
\def\1i{\'{\i}} 
\def\eq#1{ \eqno({#1}) \qquad }
\font\small=cmr7
\noindent {\bigggfnt  Simple quantum systems in the momentum representation}
\vskip 22 pt
\noindent{\bigfnt H.\ N.\ N\'u\~nez-Y\'epez}\footnote{\dag}{E-mail: nyhn@xanum.uam.mx}\par
\noindent Departamento de F\1isica, Universidad Aut\'onoma Metropolitana-Iztapalapa,  Apar\-tado Postal 55-534, Iztapalapa 09340 D.F.\ M\'exico,  \Par

\vskip 8pt

\noindent {\bigfnt E.\ Guillaum\1in-Espa\~na, R.\ P.\ 
Mart\1inez-y-Romero,\footnote{\ddag}{\rm On leave from Fac.\ de Ciencias, UNAM. E-mail: rodolfo@dirac.fciencias.unam.mx} A.\ L.\ Salas-Brito\footnote{\P}{\rm E-mail: asb@hp9000a1.uam.mx or asb@data.net.mx}}\par
\noindent{Laboratorio de Sistemas Din\'amicos, Departamento de Ciencias B\'asicas, Universidad Aut\'onoma Metropolitana-Azcapotzalco}, Apartado Postal 21-726, Coyoacan 04000 D.\ F.\ M\'exico \Par
\vskip 28 pt

\baselineskip 16 pt
\centerline{\bigfnt Abstract}\Par

The momentum representation is seldom used in quantum mechanics courses. Some students are thence surprised by the change in viewpoint when, in doing advanced work, they have to use the momentum rather than the coordinate representation. In this work, we  give an introduction to quantum mechanics in  momentum space, where the Schr\"odinger equation becomes an integral equation. To this end we discuss standard problems, namely, the free particle, the quantum motion under a constant potential, a particle interacting with a potential step, and the motion of a particle under a harmonic potential. What is not so standard is that  they are all conceived from momentum space and hence they, with the exception of the free particle, are not equivalent to the coordinate space ones with the same names. All the problems are  solved within the momentum representation making no reference to the systems they correspond to in the coordinate representation.\Par
\vfill
\eject

\noindent {\bf 1. Introduction}\Par

In quantum mechanics the spatial position, $\hat x$, and the linear momentum, $\hat p$,  operators play very symmetrical roles, as it must be obvious from the fundamental commutation relation

$$  [ \hat p, \, \hat x]=\hat p \hat x-\hat x \hat p= -i \hbar \eqno(1) \qquad$$

\noindent where, apart from a minus sign, the roles of $\hat x$ and $\hat p$ are the same. Notice that a hat over the symbols has been  used  to identify operators.  However, this fundamental symmetry is not apparent to many students of quantum mechanics because an excessive emphasis is put on the coordinate representation in lectures and in textbooks.  Some students are even lead to think of the coordinate space wave function $\psi(x)$ as more fundamental, in a certain way, than its momentum space counterpart $\phi(p)$; for, even in one of those rare cases where the Schr\"odinger equation is solved in momentum space, as is often the case with the linear potential $ x$ (Constantinescu and Magyari 1978), many students  feel that the quantum solution is somewhat not complete until the coordinate space wave function has been found.  This is a pity since a great deal of physical consequences are  understood better and many physical effects are more readily evaluated in the momentum  rather than in the coordinate representation; as an example just think of  scattering processes and form factors of every kind (Taylor 1972, Ch 3; Frauenfelder and Henley 1974;  Bransden and Joachaim 1983; Griffiths 1987). To give another interesting  example, let us remember the  one-dimensional hydrogen atom, an apparently simple system whose properties were finally understood, after  thirty years of  controversy (Loudon 1959, Elliot and Loudon 1960, Haines and Roberts 1969, Andrews 1976, Imbo and Sukhatme 1985, Boya \ea\ 1988,  N\'u\~nez-Y\'epez \ea\ 1988, 1989, Mart\'{\i}nez-y-Romero \ea\ 1989a,b,c) only after an analysis was carried out in the momentum representation (N\'u\~nez-Y\'epez \ea\ 1987, 1988, Davtyan \ea\ 1987).

  But, besides particular ocurrences, the advantages of an early introduction to the momentum representation in quantum mechanics are manyfold: a) to emphasize the basic symmetry between the representations, b) to introduce from the beginning and in a rather natural manner, distributions---pinpointing that the eigenfunctions are best regarded as generalized rather than ordinary functions---, non-local operators and integral equations, c) to help clarify  the different nature of operators in both representations, for example, in the momentum representation a free particle (vanishing potential in any representation) {\sl cannot} be considered as equivalent to a particle acted by a constant (in momentum space) potential, since this last system  admits a bound state. According to us the problems discussed in this work make clear, using rather simple examples, the distinct advantages and perhaps some of the disadvantages of working in the momentum representation. \Par

\noindent {\bf 2. Quantum mechanics in momentum space.}\Par

For calculating the basic properties and the stationary states of quantum mechanics systems, the fundamental equation is the time-independent  Scrh\"odinger equation which, in the coordinate representation, can be written as the differential eigenvalue equation

$$  -{\hbar^2 \over 2m} {d^2 \psi(x) \over dx^2} + U(x) \psi(x) = E \psi(x). \eqno(2) \qquad $$ 

\noindent This can be obtained using, in the classical Hamiltonian $H= p^2/2m +U(x)$, the operator correspondence $\hat p\to -i\hbar d/dx $, $\hat x\to x$ [operators complying with (1)]. It is equally possible the use of the alternative operator correspondence $\hat p\to p$ and $\hat x\to i \hbar   d/dp$ [also complying with (1)] that can be shown  to lead ---though not as straightforwardly as in the previous case--- to the integral Schr\"odinger equation in the momentum representation

$$ {p^2\over 2m} \phi(p) + \int dp' U(p-p') \phi (p') = E \phi(p), \eq{3} $$

\noindent where $U(p)$ is the Fourier transform of the potential energy function in the coordinate representation:

$$ U(p)= {1\over \sqrt{2\pi\hbar}}\int_{-\infty}^{+\infty}\exp(-ipx/\hbar) U(x) dx. \eq{4} $$

\noindent As it is obvious from (3), in this representation the potential energy  becomes an integral  operator (hence, usually non-local) in momentum space. Equations (2) and (3) are, in fact, Fourier transforms of each other; therefore the relationship between the coordinate and the momentum space wave functions is 

$$ \eqalign{\phi(p)=& {1\over 2\pi \hbar } \int_{-\infty}^{+\infty}\exp(-ipx/\hbar) \psi(x)\, dx, \qquad\hbox{and} \cr\cr
\psi(x)=&  \int_{-\infty}^{+\infty}\exp(+ipx/\hbar) \phi(p)\, dp.} \eq{5}  $$
   
\noindent Both functions, $\psi(x)$ and $\phi(p)$, characterize completely and independently the state of the system in question; although they differ  slightly in interpretation: whereas $\psi(x)$ is the probability amplitude that the a measurement of position gives a value in the  interval $[x,\, x+dx]$,   $\phi(p)$ is the  probability amplitude that a measurement of momentum gives a value in the interval $[p,\, p+dp]$. It should be remarked that the normalization of a function in momentum space does not guarantee the normalization of the corresponding function in coordinate space, for this to be true you should multiply the transform of $\phi(p)$ times $\sqrt{2\pi\hbar}$.

In spite of their complete equivalence, the momentum representation could throw light in certain features that may remain hidden in the coordinate representation; very  good examples of this are the SO(4) symmetry of the hydrogen atom first uncovered by Fock (1935) using his masterly treatment of the problem in the momentum representation; or  the treatment of resonant Gamow states in the momentum representation where they were found to be, contrary to what happens in the coordinate representation, square integrable solutions to a homogeneous Lippmann-Schwinger equation (Hern\'andez and Mondrag\'on 1984).

 In this work we calculate the bound energy eigenstates and corresponding quantum levels of the simplest one-dimensional potential problems in the momentum representation. We choose to present them in order of increasing  complexity, as it is usually done in basic quantum mechanics:\par

\noindent 1) The free particle with $U(p)=0$.\par
\noindent 2) Particle in a constant potential: $U(p)= -U_0$ ($U_0>0$).\par
\noindent 3) Particle interacting with the potential step

$$ U(p)= \cases {0,&  if $p >  0$,\cr\cr
                  i \alpha  \hbox{ ($\alpha$ a positive constant)},&    if $p \leq 0$;}  \eq{6} $$

 \noindent please notice that as we assume $\alpha $ to be a real number, the $i$ factor is necessary to assure the Hermiticity of the potential energy operator.\par

\noindent 4) Motion in the harmonic potential
       
$$ U(p)= -f_0 \cos(ap), \eq{7}  $$

\noindent where $f_0>0$ and $a$ are real numbers. \par

As we intend to illustrate in this contribution, in many instances the eigenfunctions are easier to calculate in momentum space than in the coordinate space representation. We have to recognize though that the momentum space eigenstates are best understood as generalized functions or distributions ---to which the  Riemann interpretation of integrals does not apply; this is explicitly illustrated by examples A, B, and D below. The energy eigenvalues are calculated, in most cases discussed here (A, B, and D), as  consistence conditions on the eigenfunctions, and in the remaining one, C, from the univaluedness of the eigenfunctions.\Par 

\noindent {\bf 3. The  examples.}\Par

We want to point out that albeit we are addressing  the same type of systems that are used to introduce quantum mechanics, here we employ the same notion of simplicity but with problems  posed in momentum space (making them very different from the coordinate space ones). Please be aware that  we use atomic units wherever it is convenient  in the rest of the paper:  $\hbar=e=m=1$. \Par

\noindent {\it A. The free particle.}\Par

In the case of the free particle, as in the coordinate representation,  $U(p)=0$ everywhere, so the Schr\"odinger equation (3) is simply

$$ \left( {p^2\over 2} -E   \right)\phi(p) =0; \eq{8} $$

\noindent this deceptively simple equation has as its basic solutions

$$  \phi_{p_E}(p)= A \delta(p-p_E)  \eq{9} $$

\noindent where $p_E$ is a solution of   $p^2={2E}$ and $A$ is a constant. This is so since, according to (8), the wave function vanishes excepting when the energy takes its  ``on shell'' value $E=p^2/2$; furthermore  as $\phi(p)$ cannot vanish everywhere, equation (9) follows. The energy eigenfunctions (9) are also simultaneously  eigenstates of the linear momentum,  

$$\hat p \phi_{p_E}(p)=p\delta(p-p_E) =p_E\phi_{p_E}, \eq{10} $$

 \noindent and form a generalized basis ---\ie\ formed by Dirac improper vectors--- for the states of a free particle  with well defined energy and linear momentum (B\"ohm 1979, Sakurai 1985); for such a free particle the most general stationary momentum-space solution is then

$$   \Phi(p)= A_+ \delta(p+|p_E|) + A_- \delta(p-|p_E|) \eq{11} $$ 

\noindent where the $A_\pm$ are  complex normalization constants; this solution represent a particle traveling to the right with momentum $|p_E|$ and to the left with momentum $-|p_E|$. The basic solutions (9) can be ``orthonormalized'' according to (Sakurai 1985)

$$   \int_{-\infty}^{+\infty} \phi^*_{p_E}(p) \phi_{p'_E}(p)\, dp = \delta(p_E-p'_E) \eq{12}             $$

\noindent which requires $A=1$ in (9). The possible energy values are constrained only by the classical dispersion relation $E=p_E^2/2m$ hence they form a continuum and the eigenstates cannot be bound.

 It is to be noted that for describing the eigenstates of a free particle, quantum mechanics uses generalized functions  for which  the probability densities $|\phi_{p_E}(p)|^2$ are not well defined! What it is well defined is their action on any square integrable function, hence on any physical state; therefore the eigenstates have to be regarded as linear functionals acting on $L^2(R)$, the set of all square integrable functions.   The only physically meaningful way of dealing with  free particles requires thus the  use of wave packets as follows 

$$\eqalign{\Phi(p)=& \int_{-\infty}^{+\infty}F(p') \,\delta(p-p') dp' \cr
                  =& F(p),} \eq{13} $$

\noindent where  $F(p)$ is any square integrable function of $p$. According to their properties then,  improper vectors, like those in (9), though very useful for formal manipulations can never strictly represent  physically realizable states (Taylor 1972, section 1a). \Par

\noindent{\it B. Motion under a constant potential}\Par

Substitution of the constant value $-U_0 <0$ into (3), gives us 

$$ \left( {p^2\over 2}-E \right)\phi(p)=U_0\int_{-\infty}^{+\infty}\phi(p') dp';  \eq{14}  $$

\noindent to solve (14), let us define the number $\check\varphi$ as

$$ \check\varphi\equiv \int_{-\infty}^{+\infty} \phi(p') dp';    \eq{15}  $$   
 
\noindent with this definition, the momentum representation Schr\"odinger equation (14) reduces to a purely algebraic equation for $\phi(p)$,

$$ \left({p^2 \over 2}-E \right) \phi(p)=U_0 \check\varphi;   \eq{16} $$

\noindent let us now define $k_0^2=-2E>0$, then  the eigenfunctions are easily seen to be

$$ \phi(p)= {2U_0\check\varphi \over p^2+k_0^2}.  \eq{17}  $$

To determine the energy eigenvalues we integrate both sides of (17) from $-\infty$ to $+\infty$  to get 

$$ \check\varphi= - 2\pi {U_0 \check\varphi\over k_0} \quad \hbox{ or } \quad E=-2\pi^2U_0^2; \eq{18}  $$

\noindent the system has a single energy eigenstate with the energy eigenvalue given in (18). The associated normalized eigenfunction is then

$$ \phi(p)= \sqrt{2\over \pi }\; {k_0^{^{3\over 2}}\over p^2 +k_0^2}.   \eq{19}$$

It is important to emphasize what we have shown: a constant potential in momentum space admits a bound state. Obviously then in this representation we have not the freedom of changing  the origin of the potential energy by adding a constant. {\sl In momentum space the potential energy is undetermined not up to a constant value but up to a Dirac-delta function potential}; that is, if you take an arbitrary potential $U(p)$ in momentum space, the physics of the problem is not changed when you consider instead the modified potential $U'(p)=U(p) + \gamma \delta(p)$ with $\gamma$ an arbitrary constant, whereas the change $U''(p)=U(p)+ \gamma\neq U'(p)$ {\bf does} indeed change the physics. The reader can prove by herself this elementary fact.  \Par

This discussion is  going forward apparently with no trouble; we have to acknowledge though  that for getting to the  condition (18), we quickly passed over a very important point, the integral of the right hand side of (17) does not exist in the ordinary Riemann sense. To obtain our result  you need  to do it instead in the distribution sense, regarding the momentum space function $\phi(p)$ as a linear functional  acting upon square integrable functions, as corresponds to  possible state functions of a quantum system. Such idea is also behind the usefulness of the delta functions as generalized basis  for the free particle states in example A.

 To particularize to the present situation, this amounts to make $\hat\varphi$ convergent  (hence meaningful) when acting on any state function (Richtmyer 1978, B\"ohm 1979). A direct way of accomplishing this is, as usually done in  theoretical physics (Taylor 1972, Frauenfelder and Henley 1974, Griffiths 1987),  to get the mentioned integral come into existence in a  principal value sense (Mathews and Walker 1970). To this end first multiply the right hand side of (17) times an $\exp(-i \epsilon\,p)$ complex factor, then perform the integral  using contour integration in the complex plane and, {\sl at the very end}, take the limit $\epsilon \to 0$. With such provisos considered, it is not difficult getting the  result (18). However, this means that the functions involved in our discussion have to be considered as linear functionals  or generalized functions, as can  be done---perhaps it would be better to say: {\sl should be done}---for every wave function of a quantum system (Messiah 1976, B\"ohm 1979); forgetting this fact can produce  erroneous results as it is exemplified by the case discussed in (N\'u\~nez-Y\'epez and Salas-Brito 1987). \Par

 It is to be noted that the free particle potential acts as a confining potential in momentum space; it allows, for each---out of a nonnegative continuum---energy value, just two choices for the momentum: $|p_E|$ and $-|p_E|$; such extreme restriction is also reflected in the wave functions, they are Dirac delta functions which peak at the just quoted values of $p$. On the other hand, the constant potential, which does not restrict the possible values of the momentum in the severe way of the zero potential,  is not as confining in momentum space and  allows a single energy eigenvalue whose associated eigenstate requires a very wide range of momenta [given in (19)] to exist. At this point we invite the reader to try to solve the problem of a particle inside an infinite potential box---{\sl in momentum space}. This is a simple and nice exercise to test the intuition on the differences between the momentum and the coordinate representation; it is not difficult to conclude that, in this case, the eigenfunctions are also Dirac delta functions  with a  lightly but subtly modified relation linking energy and momentum.\Par

\noindent{\it C. Motion in a potential step}\Par

In this case $U(p)$ is given in  (6). Using such potential, the Schr\"odinger equation becomes a simple Volterra integral equation

$$ \left({p^2\over 2} -E\right)\phi(p)+ i \alpha  \int_{-\infty}^{p} \phi(p')\, dp'. \eq{20} $$

\noindent To solve this equation, we derive both members and, using $k_0^2\equiv -2E$, we obtain a very simple differential equation

$$ {d\,\phi(p) \over dp}= 2\,{p-i\alpha\over p^2 + k_0^2}\,\phi(p),   \eq{21} $$

\noindent whose solution is

$$ \phi_{k_0}(p) = {A\over p^2 + k_0^2} \left[ {k_0-i p \over k_0 + i p} \right]^{\alpha /k_0}
\eq{22}  $$

\noindent with $A$ an integration constant.

 The energy eigenvalues follow, not from a consistency condition as in the last example, B, but from the requirement that the eigenfunctions be single valued. This  is only possible if $\alpha /k_0$ takes nonnegative integer values (Churchill 1960), \ie\ if $ k_0= {\alpha / n}$, $n=1,2,\dots$, the value $n=0$ is not allowed for $\phi(p)$ would vanish identically in that case (N\'u\~nez-Y\'epez \ea\ 1987).  Thus, the system has  an infinite number of bound energy eigenstates with energies given by

$$ E_n= -{\alpha ^2\over 2n^2},\quad n=1, 2, \dots; \eq{23} $$ 

\noindent the normalization of the eigenfunctions requires that $A=(2 \alpha ^3/n^3\pi)^{1/2}$ in equation (22). 

A very important property of the eigenfunctions is 

$$ \int_{-\infty}^{+\infty} \phi(p) dp= 0, \eq{24}  $$ 

\noindent this is required to guarantee the Hermiticity of the Hamiltonian operator of the problem (Andrews 1976, N\'u\~nez-Y\'epez \ea\ 1987, Salas-Brito 1990). We pinpoint that the  potential step in momentum space is particularly interesting because it is closely related to the study of the momentum space behaviour of electrons interacting with the surface of liquid helium, with the properties of the an hydrogen atom in superstrong magnetic fields, and with certain supersymmetric problems  (Cole and Cohen 1969,  Imbo and Sukhatme 1985, N\'u\~nez-Y\'epez \ea\ 1987, Mart\'{\i}nez-y-Romero \ea\ 1989c,   Salas-Brito 1990).\Par

\noindent {\sl D. Motion under a harmonic potential}\Par

Let us, as our final example, study the motion of a particle under the harmonic potential $U(p)= -f_0\cos(ap)$, where $a$ and $f_0 > 0$ are real constants. The Schr\"odinger equation is then

$$ {p^2\over 2}\phi(p) -f_0 \int_{-\infty}^{+\infty} \cos[a(p-p')]\phi(p') dp'=E\phi(p).\eq{25} $$

\noindent By changing $p$ for $-p$ in (25) we can show that the Hamiltonian commutes with the parity operator, thus  its eigenfunctions can be chosen as even or odd functions, \ie\ as parity eigenstates.

For solving (25), let us define $k^2\equiv -2E$ and, using the identity $2\cos x=\exp(ix)+\exp(-ix)$, we easily obtain the eigenfunctions as

$$ \phi(p)={2f_0  \over p^2+k^2} \left[\check\varphi_+ e^{-iap}+\check\varphi_-e^{+iap}  \right], \eq{26} $$

\noindent where the numbers $\check\varphi_\pm$ are defined by 

$$ \check\varphi_\pm\equiv  \int_{-\infty}^{+\infty} e^{\pm iap'}\phi(p') dp'. 
\eq{27} $$

\noindent As in  the constant potential (example B), the energy eigenvalues follow from using the definitions (27) back in the eigenfunctions (26)---please remember that we require the functions (26)   be regarded as in example B, for doing the integrals (27).  The integrals gives us the following two conditions

$$ \eqalign{\check\varphi_+=&{2f_0 \pi\over k} [\check\varphi_+ + \check\varphi_- \exp(-2ak)], \cr
    \check\varphi_-=&{2f_0 \pi\over k} [\check\varphi_- + \check\varphi_+ \exp(+2ak)]. }
\eq{28}$$

\noindent  As the eigenfunctions can be chosen as even or odd, we can select the numbers $\check\varphi_\pm $ to comply with $\check\varphi_+=\pm \check\varphi_-$, so the eigenfunctions are 

$$ \eqalign{\phi_+(p)=& {A_+\over p^2 + k^2} \cos(ap),\cr
            \phi_-(p)=& {A_-\over p^2 + k^2} \sin(ap);} \eq{29}  $$

\noindent which correspond to the complete set of eigenfunctions of the problem. 

From (28) we also get the two equations determining the energy eigenvalues

$$ {k\over  2f_0\pi} -1 =\pm  e^{-2ak}. \eq{30} $$

\noindent  one for the odd and the other for the even state.  As can be seen in Figure 1, in general equations (30) admits two solutions, let us call them $k_+$ (for the even state) and $k_-$ (for the odd state). Therefore the system has a maximum of two eigenvalues $E_+=-k_+^2/2$ and $E_-=-k_-^2/2$; the ground state is always even and the excited (odd) state exist only if $f_0\leq f_0^{\hbox{\small crit}}=1/4a\pi\simeq 0.0796$. 

 The analysis is easily done using the graph shown in Figure 1,  where we plot together $-\alpha k + 1$, $ \alpha k - 1$ and $ \exp(-2a k) $ [using $\alpha\equiv 1/(2f_0 \pi)=1$] against $k$,  for illustrating the roots of (30). In the plot, we have used  the values $a=1$, $f_0=1/2\pi\simeq 0.1592 $, corresponding to the roots $k_- = 0.7968$ (the leftmost root) and $k_+ \simeq 1.109$ (the rightmost root); thus  the energy eigenvalues are  $E_+\simeq-0.6148$ (the ground state) and $E_-\simeq-0.3175$ (the excited state). The criterion for the existence of the excited state and the value for $f_0^{\hbox{\small  crit}}$ follows from Figure 1, by noting that such critical value stems  from the equality of the slopes of the two curves meeting at the point $(0, 1)$ in the plot. Notice also that the results previously obtained for the constant potential (example B) can be recovered as a limiting case of the harmonic potential if we let $a\to 0$. \Par

\noindent{\bf 4. Concluding remarks}\Par

We have discussed four instructive one-dimensional examples in quantum mechanics from the point of view of momentum space. Purportedly we have not made any reference to the problems they represent in the coordinate representation. We expect to contribute with our approach to  the development of  physical insight for problems posed in the momentum representation and, furthermore, to help students to understand the different features of operators, as opposed to classical variables, in different representations. We also expect to made clear that sometimes it is better to treat a problem from the momentum space point of view since the solution can be simplified. The point at hand is the simple form in which the momentum space eigenfunctions are obtained in the problems discussed here; though these have to be regarded as distributions for obtaining the associated energy eigenvalues.

 With goals as the mentioned in mind and to point out other advantages of the momentum representation, in a formal set of lectures and depending on the level of the students, it may be also convenient to discuss more complex problems: as scattering and dispersion relations (Taylor 1972), or the study of resonant states as solutions of a Lippmann-Schwinger equation in momentum space (Hern\'andez and Mondrag\'on 1984),  or the 3D hydrogen atom, whose solution using Fock's method is nicely exposed in (Bransden and Joachaim 1983).\Par

Just in the case you are wondering and have not found the time for doing the transformations yourself, let us say that the problems, save the free particle, we have posed and solved in this paper are known, in the coordinate representation, as 1) the attractive delta potential (Example B), 2)  quantum motion under the (quasi-Coulomb) potential $1/x$ (Example C) and, finally,  the problem of two equal (intensity: $ A= -\pi f_0/\sqrt{2}$), symmetrically placed, attractive delta function potentials, which are displaced by $2 a$ from one another (Example D). \Par

\noindent{\bf Acknowledgements}\par

This paper was partially supported by PAPIIT-UNAM (grant IN--122498).  We  want to thank Q Chiornaya, M Sieriy, K Hryoltiy, C Srida, M Mati, Ch Cori, F Cucho,  S Mahui, R Sammi, and F C Bonito for their encouragement. ALSB also wants to thank the attendants of his UAM-A lectures on quantum mechanics (F\'{\i}sica Moderna, 99-O term), especially Arturo Vel\'azquez-Estrada and El\'{\i}as Serv\'{\i}n-Hern\'andez, whose participation was relevant for testing the ideas contained in this work. \Par

\vfill
\eject

\noindent{\bf References}\Par
\frenchspacing
\item {} Andrews M 1976 {\sl Am.\ J.\ Phys.\ } {\bf 44} 1064 \Par

\item {} B\"ohm A 1979 {\it Quantum Mechanics } (New York: Springer) \Par

\item {} Boya J, Kmiecik M, and B\"ohm A 1988 {\sl Phys.\ Rev.\ A} {\bf 37} 3567\Par

\item {} Bransden B H and Joachaim C J 1983 {\it Physics of Atoms and Molecules} (London: Longman) Ch 2\Par

\item {} Cole M W and Cohen M H 1969 {\sl Phys.\ Rev.\ Lett.\ } {\bf 23} 1238\Par

\item {} Churchill R V 1960 {\it Complex Variables and Applications} (New York: McGraw-Hill) pp 59--60\Par

\item {} Constantinescu F and Magyari E 1978 {\it Problems in Quantum Mechanics} (London: Pergamon) Ch V problem 118\Par

\item {} Davtyan L S, Pogosian G S, Sissakian A N and Ter-Antonyan V M 1987 {\sl J.\  Phys.\ A: Math.\ Gen.\ } {\bf 20} 2765\Par

\item {} Elliot R J  and Loudon R 1960 {\it J.\ Phys.\ Chem.\ Solids} {\bf 15} 196\Par

\item {} Fock V A 1935 {\sl Z.\ Phys.\ } {\bf 98} 145\Par

\item {} Frauenfelder H and Henley E M 1974 {\it Subatomic Physics} (New Jersey: Pren\-tice-Hall) Ch 6 \Par

\item {} Griffiths D 1987 {\it Elementary Particles} (Singapore: Wiley) Ch 8\Par

\item  {} Haines L K and Roberts D H 1969 {\sl Am.\ J.\ Phys.\ } {\bf 37} 1145\Par

\item {} Hern\'andez E and Mondrag\'on A 1984 {\it Phys.\ Rev.\ C } {\bf 29} 722\Par

\item {} Imbo T D and Sukhatme U P 1985 {\sl Phys.\ Rev.\ Lett.\ } {\bf 54} 2184\Par

\item {} Loudon R 1959 {\sl Am.\ J.\ Phys.\ } {\bf 27} 649\Par

\item {}  Mart\'{\i}nez-y-Romero R P, N\'u\~nez-Y\'epez H N, Vargas C A and Salas-Brito A L 1989a {\sl Rev.\ Mex.\ Fis.\ } {\bf 35}  617. \Par 

\item {}  Mart\'{\i}nez-y-Romero R P, N\'u\~nez-Y\'epez H N, Vargas C A and Salas-Brito A L 1989b {\sl Phys.\ Rev.\  A} {\bf 39}  4306. \Par 
 
\item {} Mart\'{\i}nez-y-Romero R P, N\'u\~nez-Y\'epez H N and Salas-Brito A L 1989c {\sl Phys.\  Lett.\ A} {\bf 142}  318\Par

\item {} Messiah A  1976 {\it Quantum Mechanics} Vol I (Amsterdam: North Holland)  p 463
\Par

\item {} Mathews J and Walker R L 1970 {\it Mathematical Methods of Physics} (New York: Benjamin) Ch 11 and Appendix 2\Par

\item {} N\'u\~nez-Y\'epez H N and Salas-Brito A L 1987 {\sl Eur.\ J.\ Phys.\  } {\bf 8} 307\Par

\item {} N\'u\~nez-Y\'epez H N, Vargas C A and Salas-Brito A L 1987 {\sl Eur.\ J.\ Phys.\  } {\bf 8} 189\Par

\item {} N\'u\~nez-Y\'epez H N, Vargas C A and Salas-Brito A L 1988 {\sl J.\  Phys.\ A: Math.\ Gen.\ } {\bf 21} L651\Par

\item {} N\'u\~nez-Y\'epez H N, Vargas C A and Salas-Brito A L 1989 {\sl  Phys.\ Rev.\ A} {\bf 39} 4306\Par

\item {} Richtmyer R D 1978 {\it Principles of Advanced Mathematical Physics } Vol I (New York: Springer)  Ch 2\Par

\item {}  Sakurai J J 1985 {\it Modern Quantum Mechanics} (Reading: Addison-Wesley) \Par

\item {} Salas-Brito A L 1990 {\it \'Atomo de Hidr\'ogeno en un Campo Magn\'etico Infinito: Un Modelo con Regla de Superselecci\'on}, { Tesis Doctoral},  Facultad de Ciencias, Universidad Nacional Aut\'onoma de M\'exico (in Spanish)\Par

\item {} Taylor J R 1972 {\it Scattering Theory} (New York: Wiley)\Par

\nonfrenchspacing

\vfill
\eject

\centerline{Figure Caption}

\noindent {\bf Figure 1}\Par

\noindent The figure illustrates the solution to equations (30) determining the energy eigenvalues under  the harmonic potential (7). We here plot $\exp(-2ak)$, $\alpha k-1$ and $-\alpha k +1$ against $k$ all in the same graph. Just for illustration purposes, we have  used the specific values $a=1$, $f_0=1/2\alpha\pi\simeq 0.1592 $,  (we have defined $\alpha= 1/2f_0\pi$ and used $\alpha=1)$. The critical value of $f_0$, giving birth to the excited state, is $f_0^{\hbox{\small crit}}=(4 a\pi)^{-1}$ as can be obtained from the equality of the slopes of the two curves meeting at the point $(0, 1)$ in the graph. In the situation exemplified by this figure, $f_0^{\hbox{\small crit}}\simeq 0.0796$ and the roots of equations (30) are $k_+\simeq 1.109$ and $k_-\simeq 0.7968$. 
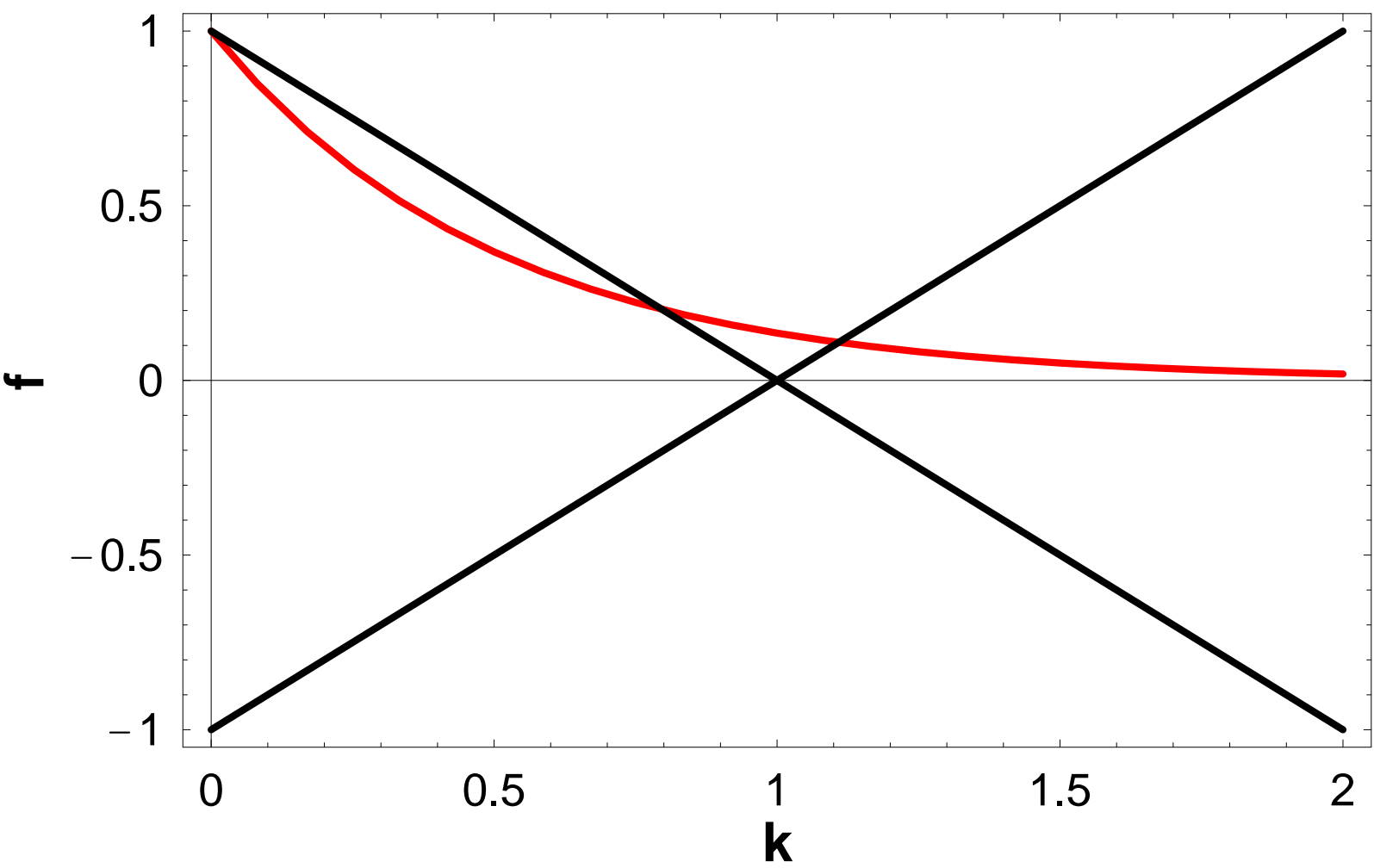
\bye